\documentclass[aps,prl,twocolumn,superscriptaddress,showpacs]{revtex4-1}

\usepackage[colorlinks,linkcolor=blue,citecolor=blue,urlcolor=blue,breaklinks=true]{hyperref}
\usepackage{amsmath,amssymb,graphics,epsfig,epstopdf,color,verbatim,braket,tabularx,breakurl}

\begin{document}

\title{Characterizing Many-Body Localization by Out-of-Time-Ordered Correlation}

\author{Rong-Qiang He}
\email{rqhe@ruc.edu.cn}
\affiliation{Institute for Advanced Study, Tsinghua University, Beijing 100084, China}
\affiliation{Department of Physics, Renmin University of China, Beijing 100872, China}
\author{Zhong-Yi Lu}
\email{zlu@ruc.edu.cn}
\affiliation{Department of Physics, Renmin University of China, Beijing 100872, China}

\date{\today}

\begin{abstract}
  The out-of-time-ordered (OTO) correlation is a key quantity for quantifying quantum chaoticity and has been recently used in the investigation of quantum holography. Here we use it to study and characterize many-body localization (MBL). We find that a long-time logarithmic variation of the OTO correlation occurs in the MBL phase but is absent in the Anderson localized and ergodic phases. We extract a localization length in the MBL phase, which depends logarithmically on interaction and diverges at a critical interaction. Furthermore, the infinite-time `thermal' fluctuation of the OTO correlation is zero (finite) in the ergodic (MBL) phase and thus can be considered as an order parameter for the ergodic-MBL transition, through which the transition can be identified and characterized. Specifically, the critical point and the related critical exponents can be calculated.

  %The second is longer than the first, thus interaction tends to delocalize the systems. The strength of this tendency increases linearly as interaction increases.
\end{abstract}


\maketitle

Recently, the proposal and studies of a simple quantum-mechanical model, known as the Sachdev-Ye-Kitaev (SYK) model \cite{noteKitaev2015,Polchinski16,Maldacena2016}, show that it has some interesting features shared by a black hole, and it is proposed to be a model of quantum holography. A key quantity in this context is the out-of-time-ordered (OTO) correlation \cite{Larkin69,Kitaev2014}, which is generalized from a quantity describing classical chaos and can be used to diagnose quantum chaos as well as scrambling of quantum information in black holes. These attract a lot of attention and there are also experimental proposals to simulate the SYK model \cite{Danshita2016} and measure this quantity in cold atom systems \cite{Swingle2016} or via a `quantum clock' \cite{Zhu2016}.

The SYK model is a disordered model in zero dimension. The disorder plays a key role in making the model most chaotic. However, when disorder is introduced in finite-dimensional models, one usually finds Anderson localization \cite{Anderson58} or many-body localization (MBL) \cite{Basko06,Gornyi05,Huse15AR,Imbrie16,Imbrie16JSP,Choi2016} if the system is weakly interacting. An MBL system is effectively integrable because of the emergence of a complete set of local integrals of motion \cite{Serbyn13,Huse14,Imbrie16,Imbrie16JSP,He16}, thus any chaoticity is suppressed and ergodicity is broken down as well. This makes MBL special. Consequently, it is difficult to describe intriguing properties of MBL by conventional correlation functions and convectional methods, especially the critical behavior of the ergodic-MBL transition and the subtle distinction between Anderson localization and MBL.

In this paper, we employ the OTO correlation to study and characterize MBL first via phenomenological analysis and then by numerical calculations for a one-dimensional interacting spinless fermionic model. We find that the OTO correlation initially decreases polynomially in short time, then it reaches zero in an ergodic phase but remains finite in the noninteracting Anderson localized phase. What is interesting is that in an MBL phase the OTO correlation decreases logarithmically to zero in long time, reminiscent of the behavior of the logarithmic increase in time of the entanglement entropy \cite{Znidaric08,Bardarson12,Serbyn13slow,Kim14}. A localization length for the MBL can be further extracted, which depends logarithmically on the interaction, diverges at a critical interaction, and hence predicts a transition---the ergodic-MBL transition. Furthermore, we find that the `thermal' fluctuation of the OTO correlation at infinite time is zero for an ergodic phase but finite for an MBL phase. Thus this fluctuation can be used as an order parameter to characterize the ergodic-MBL transition, of which the critical point and related critical exponents can be calculated, for example.

{\em Model.}---For the sake of concreteness, a one-dimensional spinless fermionic model with nearest-neighbor hoppings, nearest-neighbor density-density interactions, and disordered on-site potentials is studied. The Hamiltonian is
\begin{equation}\label{eq:model}
\hat{H} = -\frac{1}{2} \sum_{\langle ij\rangle} (c_{i}^{\dagger}c_{j}+c_{j}^{\dagger}c_{i}) + V \sum_{\langle ij\rangle} (\hat{n}_{i} - \frac{1}{2}) (\hat{n}_{j} - \frac{1}{2}) - \sum_{i} \mu_i \hat{n}_i,
\end{equation}
where $\mu_i$ is randomly chosen in $[-w, w]$ with uniform distribution and $\hat n_i = c_i^\dagger c_i$. $L$ is the number of lattice sites. A periodic boundary condition is chosen. This model can be transformed into a random-field $XXZ$ spin chain by the Jordan-Wigner transformation. For $V > 0$, there is an ergodic to MBL phase transition as $w$ increases, while the system is Anderson localized for $V = 0$ and $w > 0$.

%$C(t) = \frac{1}{2} \langle \hat V^\dagger(0) \hat W^\dagger(t) \hat W(t) \hat V(0) \rangle + \frac{1}{2} \langle  \hat W^\dagger(t) \hat V^\dagger(0) \hat V(0) \hat W(t) \rangle + \Re F(t)$.

{\em OTO correlation.}---The OTO correlation for two commuting/anticommuting operators $\hat W$ and $\hat V$ is defined as
\begin{equation}\label{eq:otocorr}
F(t) = \pm \langle  \hat W^\dagger(t) \hat V^\dagger(0) \hat W(t) \hat V(0) \rangle,
\end{equation}
where $+$ ($-$) is chosen when $\hat W$ and $\hat V$ commute (anticommute) and $\hat W(t) = \exp(iHt) \hat W \exp(-iHt)$ with $\hbar = 1$, and the average $\langle \dotsm \rangle$ is taken on some ensemble described by a density operator $\hat\rho$, i.e., $\langle  \hat O \rangle = {\rm tr} \hat\rho \hat O$. The OTO correlation arises from a commutator square
\begin{equation}\label{eq:otocomm}
C(t) = \frac{1}{2} \langle [\hat W(t), \hat V(0)]_\pm^\dagger [\hat W(t), \hat V(0)]_\pm \rangle,
\end{equation}
which is non-negative and is zero at $t = 0$. Let us call it `OTO commutativity' for convenience. Usually, $\hat W$ and $\hat V$ are chosen to be local operators in different locations. $\hat W(t)$ spreads in space as time. $C(t)$ becomes significant when the spread of $\hat W(t)$ reaches the location of $\hat V$. When $\hat W$ and $\hat V$ are unitary, $C(t) = 1 - \Re F(t)$, and this is the case on which we focus below. For clarity of later references, when choosing $\hat W = \kappa_i$ and $\hat V = \kappa_j$ with $\kappa_i$ being an operator at site $i$, we denote the OTO correlation/commutativity as $F_{ij}(\kappa; t)$/$C_{ij}(\kappa; t)$, and further denote as $F_{ij}^{(n)}(\kappa; t)$/$C_{ij}^{(n)}(\kappa; t)$ when choosing $\hat\rho = | n \rangle \langle n |$ with $| n \rangle$ being an eigenstate of the system.

%$F_{ij}(\eta;t) = - \langle  \eta_i(t) \eta_j(0) \eta_i(t) \eta_j(0) \rangle$
% if we choose $\hat W$ and $\hat V$ to be the Majorana fermion operators $\eta_i$ and $\eta_j$ respectively, i.e.

{\em Phenomenological analysis.}---In the MBL phase, the Hamiltonian (\ref{eq:model}) can be transformed into a simple form in terms of a complete set of local integrals of motion $\{ \hat\tau_i \}$ ($[\hat\tau_i, \hat\tau_j] = [\hat\tau_i, \hat H] = 0$) \cite{Serbyn13,Huse14}:
\begin{equation}\label{eq:liommodel}
\hat{H} = \sum_i \xi_i \hat\tau_i + \sum_{ij} V_{ij} \hat\tau_i \hat\tau_j + \sum_{ijk} V_{ijk} \hat\tau_i \hat\tau_j \hat\tau_k + \cdots,
\end{equation}
where $\xi_i, V_{ij}, V_{ijk}, \dotsc$ are coupling coefficients and $V$'s decays exponentially as the distance between $i, j, k, \dotsc$ increases. The locality of $\hat\tau_i$ is manifested by $\hat\tau_i = \hat U \hat n_i \hat U^\dagger$, where $\hat U$ is a local unitary operator and diagonalizes the Hamiltonian, as shown in Ref.~\cite{He16}. The OTO correlation $F_{ij}(\eta;t)$ can be formally obtained for this model, where $\eta_i \equiv \hat U \gamma_i \hat U^\dagger$ with $\gamma_i \equiv c_i + c_i^\dagger$ being a Majorana fermion operator. Note that $\eta_i$ and $\gamma_i$ are unitary and Hermitian.
\begin{eqnarray}\label{eq:Ftau}
F_{ij}(\eta; t) &=& \sum_n \rho_n F_{ij}^{(n)}(\eta; t) \\
          &=& \sum_n \rho_n \exp[i t \Delta\tau_i^{(n)} \Delta\tau_j^{(n)} \tilde V_{ij}^{(n)}] \\
          &=& \int_{-\infty}^{\infty} f(x) \exp(i t x) dx,
\end{eqnarray}
where $\hat\rho =  \sum_n \rho_n | n \rangle \langle n |$ describes an ensemble and $ | n \rangle$ is an eigenstate of $\hat H$. $F_{ij}^{(n)}(\eta; t) = \exp[i t \Delta\tau_i^{(n)} \Delta\tau_j^{(n)} \tilde V_{ij}^{(n)}]$. $\tau_i^{(n)} \in \{0, 1 \}$ is the eigenvalue of $\hat\tau_i$ on $| n \rangle$.  $| n \backslash i \rangle \equiv \eta_i |n\rangle$ is also an eigenstate of $\hat H$ and $\Delta\tau_i^{(n)} \equiv \tau_i^{(n \backslash i)} - \tau_i^{(n)}$ with $|\Delta\tau_i^{(n)}| = 1$. $\tilde V_{ij}^{(n)} = V_{ij} + \sum_k V_{ijk} \tau_k^{(n)} + \cdots $ is an effective coupling strength between $\hat\tau_i$ and $\hat\tau_j$ for state $|n\rangle$, which is bounded and decays exponentially as $r \equiv |j - i|$ increases, i.e., $\tilde V_{ij}^{(n)} \sim V_r \equiv V \exp(-r/\xi)$, where $\xi$ defines a localization length.  $f(x) = \sum_n \rho_n \delta(x - \Delta\tau_i^{(n)} \Delta\tau_j^{(n)} \tilde V_{ij}^{(n)})$ can be considered as a probability density function with a standard deviation $\sim V_r$.

%$\hat H | n \rangle = E_n | n \rangle$
%$F_{ij}(\eta;t) = \sin(t V_r) / t V_r$ for $f(x) = \Theta (V_r - x) / 2 V_r$.
%\begin{eqnarray}
%b V_r t \ll 1,  & \quad & C_{ij}(\eta; t) \sim (b V_r t)^\alpha; \label{eq:Ct1} \\
%b V_r t \sim 1, & \quad & C_{ij}(\eta; t) \sim 1-e^{-1} + e^{-1}\alpha \ln b V_r t; \label{eq:Ct2} \\
%b V_r t \gg 1,  & \quad & C_{ij}(\eta; t) \sim 1. \label{eq:Ct3}
%\end{eqnarray}

In the thermodynamic limit, $f(x)$ is a continuous function so that $F_{ij}(\eta;t)$ decreases from 1 to 0 as time goes from 0 to $\infty$. For example, $F_{ij}(\eta;t) = \exp(- V_r^2 t^2 / 2)$ for $f(x) \propto \exp( - x^2 / 2 V_r^2)$ being a normal distribution and $F_{ij}(\eta;t) = \exp(- V_r t)$ for $f(x) = V_r / \pi (x^2 + V_r^2)$ being a Lorentz distribution. Generally, $F_{ij}(\eta;t)$ may be modeled by $F_{ij}(\eta;t) = \exp[-(b V_r t)^\alpha]$ with $b \sim 1$ and $\alpha > 0$ being parameters, which is surprisingly good as demonstrated by numerical results below. Accordingly, the corresponding OTO commutativity is
\begin{equation}\label{eq:Cmodel}
C_{ij}(\eta;t) = 1 - \exp[-(b V_r t)^\alpha].
\end{equation}
For small $t$, $C_{ij}(\eta; t)$ increases polynomially as time, which is essentially different from the exponential increase of $C_{ij}(\eta;t)$ for a quantum chaotic system. More interestingly,
\begin{equation}\label{eq:Clog}
C_{ij}(\eta; t) \approx 1-e^{-1} + e^{-1}\alpha \ln b V_r t
\end{equation}
for $b V_r t \sim 1$, i.e., at intermediate time $C_{ij}(\eta; t)$ increases logarithmically as time. This resembles the logarithmic growth of entanglement entropy \cite{Znidaric08,Bardarson12,Serbyn13slow,Kim14}. Actually, they share the same origin of the interaction induced dephasing. For $C_{ij}(\eta; t)$ becoming significant, say $C_{ij}(\eta; t) = 1-e^{-1}$, one obtains $r = \xi \ln b V t$, implying that $\hat W(t)$, carrying some information, spreads logarithmically slowly in space.

In the Anderson localization case ($V = 0$), one finds immediately that $\tilde V_{ij}^{(n)} = 0$ \cite{Serbyn13,Huse14,He16} so that $F_{ij}(\eta;t) = 1$ and $C_{ij}(\eta;t) = 0$, which is in sharp contrast with the MBL case where $C_{ij}(\eta;t)$ increases from 0 and saturates finally to 1 as time increases. So the OTO correlation can be used to distinguish MBL from Anderson localization. A conventional time-ordered correlation function, such as $\langle \eta_i(t) \eta_j(t) \eta_i(0) \eta_j(0) \rangle$ or $\langle \eta_i(0) \eta_i(t) \eta_j(t) \eta_j(0) \rangle$, lacks this feature because contributions from the one-body term $\sum_i \xi_i \hat\tau_i$ in the Hamiltonian (\ref{eq:liommodel}) cannot be canceled out. Refer to Refs.~\cite{Serbyn2014quenches,Serbyn2014interferometric,Vasseur2015} for related discussions.

The imperfection of this phenomenological analysis is the choice of the two operators $\hat W$ and $\hat V$ in the OTO correlation. They are chosen to be $\eta_i$ and $\eta_j$ above. $\eta_i$ is quasilocalized in the MBL phase, but becomes extended when the system enters the ergodic phase. This qualitative change of $\eta_i$ across the transition hides in some extent the singular behavior of the transition from the OTO correlation. It may thus be difficult to identify and characterize the transition with this choice of the OTO correlation. A resolution to this difficulty is a new choice of the OTO correlation with $\hat W = \gamma_i$ and $\hat V = \gamma_j$ totally localized, namely, $F_{ij}(\gamma;t)$, which is the same choice as that in the studies of the Sachdev-Ye-Kitaev model. The cost is that now it is difficult to find the properties of the OTO correlation analytically. So we resort to a numerical calculation, as presented below.

%$\rho_n = {\rm constant}$ for eigenstates with eigenenergies in a small energy window and zero outside. The position of the energy window is denoted by $\epsilon = (E - E_{\rm min}) / (E_{\rm max} - E_{\rm min})$. $\epsilon = 0.5$, namely the center of the energy spectrum.

{\em Numerical results.}---We use the microcanonical ensemble in the calculation. A relative energy $\epsilon \equiv (E - E_{\rm min}) / (E_{\rm max} - E_{\rm min})$ is introduced for convenience, where $E$ is the energy of the system and $E_{\rm max}$ and $E_{\rm min}$ are the maximal and minimal energies of the many-body eigenenergy spectrum for a specific disorder realization. $\epsilon = 0.5$ is set in the following calculation, which corresponds to infinite temperature for a canonical ensemble and is most relevant for MBL and the ergodic-MBL transition. In the calculation, the OTO correlation is averaged over a number of ($\sim 10^4$) independent disorder realizations. After averaging, $ij$-subscripted quantities [e.g., $C_{ij}(\gamma;t)$] depend only on $r = |j - i|$ and we replace the subscript $ij$ with $r$ [e.g., $C_r(\gamma;t)$].

%---------------------------------------------------------------------Fig.1---------------------------------------------------------------------------------------------------------------------
\begin{figure}[tb]
  \includegraphics[width=\columnwidth]{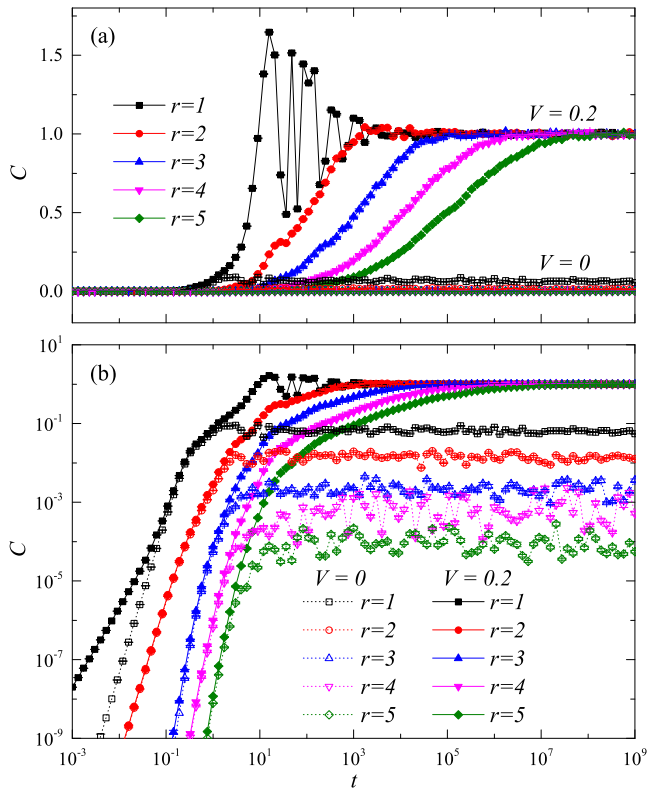}
  \caption{\label{fig:Vcmp} (Color online) The disorder averaged OTO commutativity $C_r(\gamma;t)$ for $V = 0$ and $0.2$ in linear scale (a) and logarithmic scale (b). $w = 8$, $\epsilon = 0.5$, and $L = 14$. For small $t$, $C$ increases polynomially as time and has no apparent difference for $V = 0$ and $V > 0$ (except the $r = 1$ case). Then, $C$ increases logarithmically as time for $V > 0$, while it saturates for $V = 0$. Finally, $C$ for $V > 0$ saturates to 1, corresponding to 0 for the OTO correlator $F_r(\gamma;t)$. }
\end{figure}
%-----------------------------------------------------------------------------------------------------------------------------------------------------------------------------------------------

A comparison for the OTO commutativity $C_r(\gamma;t)$ between the Anderson localized ($V = 0$) and the MBL ($V > 0$ but small) phases is shown in Fig.~\ref{fig:Vcmp}. In the two cases, $C_r(\gamma;t)$ shares nearly the same values (except for the most beginning stage for $r = 1$) and increases polynomially for small $t$ [see Fig.~\ref{fig:Vcmp}(b)]. In this stage, the interaction does not play a role, as we can see. The action of $\gamma_i$ or $\gamma_j$ on an eigenstate will generate a state composed of a number of eigenstates. The subsequent short-time evolution, being a local relaxation, is determined by the one-body energies $\xi_i \sim w$ in short time $t < w^{-1}$. The interaction energies $\sim V_r \equiv V \exp(-r/\xi)$ are small and have no effect for small $t$. At $t \sim w^{-1}$ the perturbation of $\gamma_i$ or $\gamma_j$ is locally fully relaxed.

%---------------------------------------------------------------------Fig.2---------------------------------------------------------------------------------------------------------------------
\begin{figure}[tb]
  \includegraphics[width=\columnwidth]{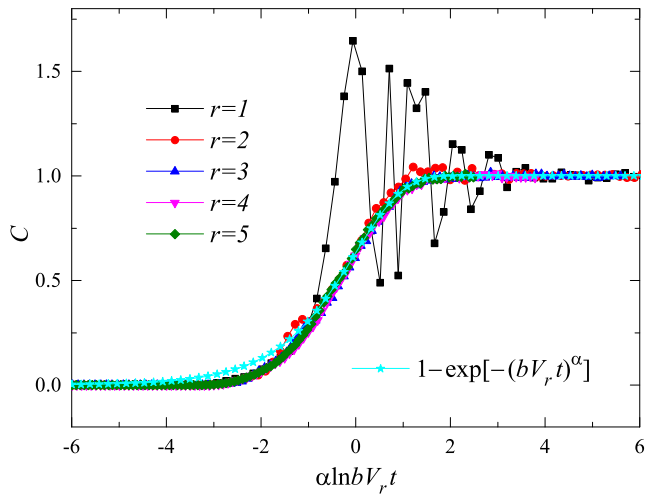}
  \caption{\label{fig:C.xi} (Color online) Data collapse for the disorder averaged OTO commutativity $C_r(\gamma;t)$ for different $r$, which can be approximated by $1 - \exp[-(b V_r t)^\alpha]$ from Eq.~(\ref{eq:Cmodel}) with $V_r \equiv V \exp(-r / \xi)$. $w = 8$, $V = 0.2$, $\epsilon = 0.5$, and $L = 14$. Fit parameters: $\xi = 0.406$ defines a localization length, $\alpha < 1$ decreases as $r$ increases, and $b = 3.41$. }
\end{figure}
%-----------------------------------------------------------------------------------------------------------------------------------------------------------------------------------------------

For $t > w^{-1}$, there is not any essential change in $C_r(\gamma;t)$ for the noninteracting case [see Fig.~\ref{fig:Vcmp}(b)]. In contrast, for the MBL case, the effective two-body interaction energy between sites $i$ and $j$, $\sim V_r$, will cause dephasing in the time evolution of many-body eigenstates as $t$ increases and approaches $V_r^{-1} = V^{-1} \exp(r/\xi)$, i.e., the exponentially small effective interaction will cause a relaxation in an exponentially long time. Thus we see a logarithmic increase as time in $C_r(\gamma;t)$ for $t \sim V_r^{-1}$ and then a saturation when $t > V_r^{-1}$, consistent with Eqs.~(\ref{eq:Cmodel}) and (\ref{eq:Clog}). A localization length $\xi$ can be extracted by a data collapse with $C_r(\gamma;t)$ guided by Eq.~(\ref{eq:Cmodel}) as we do in Fig.~\ref{fig:C.xi}. The model for the OTO commutativity (\ref{eq:Cmodel}) matches excellently with the numerical data, as shown in Fig.~\ref{fig:C.xi}, except for a little deviation at and before the onset of the logarithmic increase. This deviation may be accounted for by the additional short-time local relaxation in $C_r(\gamma;t)$ rather than in $C_r(\eta;t)$.

%---------------------------------------------------------------------Fig.3---------------------------------------------------------------------------------------------------------------------
\begin{figure}[tb]
  \includegraphics[width=\columnwidth]{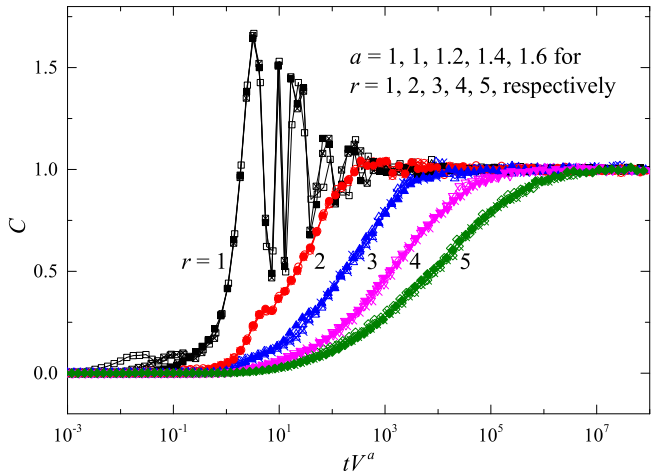}
  \caption{\label{fig:C.tVa} (Color online) Data collapse for the disorder averaged OTO commutativity $C_r(\gamma;t)$ for $V = 0.01$ (open), $0.05$ (open with cross), and $0.20$ (solid) in the MBL phase. $w = 8$, $\epsilon = 0.5$, and $L = 14$. The horizontal axis is rescaled, where $a = 1$ for $r = 1$ and $a = 1 + 0.2 (r - 2)$ for $r = 2, \dotsc, 5$. }
\end{figure}
%-----------------------------------------------------------------------------------------------------------------------------------------------------------------------------------------------

To investigate the effect of the interaction more carefully, we have calculated $C_r(\gamma;t)$ for several different interactions in the MBL phase. A good data collapse is shown in Fig.~\ref{fig:C.tVa} when the horizontal axis $t$ is rescaled as $t V^a$ with $a = 1 + u (r - 2)$ (where $u > 0$ is a constant) for $r \ge 2$, implying that for the long-range cases the effective two-body interactions $\tilde V_{ij}^{(n)} \sim V^a \exp[-(r-2)/\xi_0] = V \exp[-(r-2)/\xi]$ with
\begin{equation}\label{eq:xiV}
\xi^{-1} = \xi_0^{-1} - u \ln V
\end{equation}
and $\xi_0$ independent of $V$. Therefore, we find that the localization length $\xi$ depends logarithmically on $V$. As expected, $\xi$ increases as $V$ increases. Remarkably, $\xi$ will diverge as $V$ approaches $V_c \equiv \exp(u^{-1} \xi_0^{-1})$, thus an ergodic-MBL phase transition is predicted. As $\xi$ goes to infinity, the time range for the logarithmic increase of $C_r(\gamma;t)$ will shrink to zero, which can be readily inferred from Eq.~(\ref{eq:Cmodel}). In the ergodic phase, $C_r(\gamma;t)$ increases polynomially fast to 1 in short time as expected (data not shown).

%For the short-range cases with $r = 1$ and $2$, the data collapse is especially good with $a = 1$, which is consistent with the analysis above very well as the effective two-body interaction ($\tilde V_{ij}^{(n)}$) between sites $i$ and $j$, which results in dephasing and determines the logarithmic increase of $C_r(\gamma;t)$, is expected to be proportional to $V$. However, this expectation has to be corrected slightly for the long-range cases with $r > 2$, where the power of $V$ is modified slightly to be $a = 1 + 0.2 (r - 2)$, implying that $\tilde V_{ij}^{(n)} \sim V^a \exp[-(r-2)/\xi_0] = V \exp[-(r-2)/\xi]$ with $\xi^{-1} = \xi_0^{-1} - 0.2 \ln V$ and $\xi_0$ independent of $V$, i.e., the localization length $\xi$ depends on $V$.

%$\tilde V_{ij}^{(n)} \sim V^a \exp(-r/\xi_0)$.
%It is worth noting that this data collapse implies that the localization length $\xi$ is independent of $V$ when $V \ll w$, which is also be shown in our previous study with local integrals of motion \cite{He16}.

%As shown in Eq.~(\ref{eq:Ftau}), $F_{ij}^{(n)}(\eta; t) = \exp[i t \Delta\tau_i^{(n)} \Delta\tau_j^{(n)} \tilde V_{ij}^{(n)}]$ will oscillate forever while the decay of $F_{ij}(\eta; t)$ results from the ensemble average, i.e., there is a `thermal' fluctuation for $F_{ij}(\eta; t)$

\begin{figure}[tb]
  \includegraphics[width=\columnwidth]{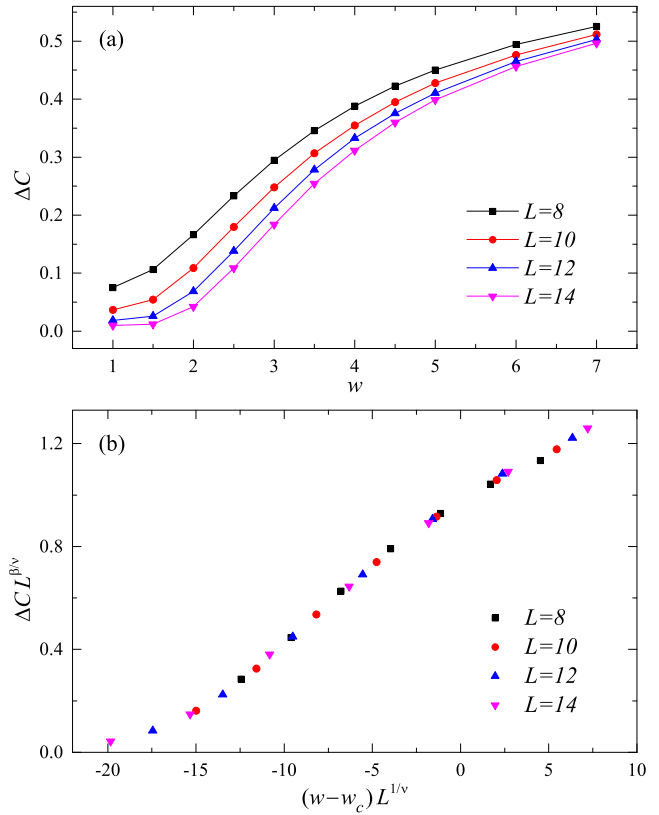}
  \caption{\label{fig:dC.r2} (Color online) (a) Disordered averaged thermal fluctuation $\Delta C_r(\gamma; t)$ [$\propto \Delta F_r(\gamma; t)$] at $t = \infty$, serving as an order parameter for the transition from the ergodic phase (small $w$) to the MBL phase (large $w$), vanishes in the ergodic phase, while it remains finite in the MBL phase as $L \rightarrow \infty$, which is more evidently shown in (b) by the data collapse. $V = 1$ and $\epsilon = 0.5$. $\Delta C_r(\gamma; \infty) \sim |w - w_c|^\beta$ for $L \rightarrow \infty$ with $w_c = 3.7$ and $\beta = 0.57$. $\nu = 1.2$. The $r = 2$ case is shown, and $\Delta C_r(\gamma; \infty)$ hardly depends on $r$. }
\end{figure}

{\em The ergodic-MBL transition.}---A more careful inspection on the OTO correlation allows us to identify and characterize the ergodic-MBL transition. We define a `thermal' fluctuation for $F_{ij}(\kappa; t)$ as
\begin{equation}\label{eq:dF}
\Delta F_{ij}(\kappa; t) = \sqrt{\sum_n \rho_n |F_{ij}^{(n)}(\kappa; t) - F_{ij}(\kappa; t)| ^ 2}.
\end{equation}
$\Delta F_{ij}(\eta; \infty) = 1$ because $|F_{ij}^{(n)}(\eta; t)| = 1$ and $F_{ij}(\eta; \infty) = 0$. But $\Delta F_{ij}(\gamma; t) < 1$ because the action of $\gamma_i$ on an eigenstate $| n \rangle$ will generate a state composed of a number of [denote this number as $N^{(n)}(\gamma_i)$] eigenstates [denoted as $\{|\psi_k^{(n)}(\gamma_i)\rangle\}$ with $k = 1, 2, \dotsc, N^{(n)}(\gamma_i)$] and then the time evolution will result in dephasing and make $|F_{ij}^{(n)}(\gamma; t)| < 1$. Note that $\hat U^\dagger \eta_i \hat U \equiv \gamma_i$ and $\hat U$ is a local unitary transformation in the MBL phase, so $N^{(n)}(\gamma_i)$ is finite and $| \psi_k^{(n)}(\gamma_i) \rangle$ (for different $k$) differ from each other only locally within a localization length \cite{He16}, resulting in a limited dephasing and hence a lower but finite thermal fluctuation for $F_{ij}(\gamma; t)$. In contrast, for the ergodic phase $\hat U$ is a global transformation so that $N^{(n)}(\gamma_i) = \infty$ and the dephasing will result in a complete destructive interference and hence $F_{ij}^{(n)}(\gamma; t) = 0$ at a sufficiently long time and $\Delta F_{ij}(\gamma; \infty) = 0$. This analysis is supported by the numerical result shown in Fig.~\ref{fig:dC.r2}.

%Note that the thermal fluctuation for $C_{ij}(\gamma; \infty)$, denoted as $\Delta C_{ij}(\gamma; \infty)$, is proportional to $\Delta F_{ij}(\gamma; \infty)$ and only the disorder averaged $\Delta C_{ij}(\gamma; \infty)$ is presented here.

%---------------------------------------------------------------------Fig.4---------------------------------------------------------------------------------------------------------------------
%-----------------------------------------------------------------------------------------------------------------------------------------------------------------------------------------------

For a fixed $V$, the system is in the ergodic phase for $w < w_c$ and enters the MBL phase for $w > w_c$. As shown in Fig.~\ref{fig:dC.r2}(a), the fluctuation of $C_{ij}(\gamma; \infty)$ approaches zero as the system size $L$ increases for small $w$ and remains finite for large $w$. A careful scaling analysis in Fig.~\ref{fig:dC.r2}(b) yields $w_c = 3.7$ for $V = 1$, being well consistent with other results in the literature. Effectively, this fluctuation can be taken as an order parameter for the ergodic-MBL transition and the related critical exponents can be calculated. Furthermore, one may find the mobility edge of the system with this quantity simply by varying $\epsilon$ \cite{Luitz15}.

%It has been known that an arbitrarily small interaction can result in a logarithmic grow of entanglement entropy in time \cite{Serbyn13slow}.

{\em Conclusion.}---In summary, we have used the out-of-time-ordered (OTO) correlation/commutativity, a key quantity for the description of quantum chaoticity and quantum holography, to study and characterize many-body localization (MBL). We find a short-time polynomial increase and long-time logarithmic increase of the OTO commutativity in the MBL phase but an absence of a long-time logarithmic increase in the Anderson localized and ergodic phases. The saturate value of the OTO commutativity can be reached finally in the MBL and ergodic phases but not in the Anderson localized phase. A localization length can be extracted in the MBL phase, which depends logarithmically on the interaction, diverges at a critical interaction, and predicts a transition---the ergodic-MBL transition. Moreover, the `thermal' fluctuation of the OTO correlation at infinite time is zero (finite) for the ergodic (MBL) phase, and thus can be considered as an order parameter for the ergodic-MBL transition and can be used to identify and characterize the transition, of which the critical point and related critical exponents can be calculated.

%The time evolution of the OTO commutativity has three stages including polynomial-increase at short time, then logarithmic-increase at long time, and finally saturation.

{\em Note added.} Recently, we became aware of a few related works \cite{Huang2016,Fan16,Chen2016,Swingle2016Slow}.

\begin{acknowledgments}
This work was supported by National Natural Science Foundation of China (Grants No. 11474356 and No. 91421304).  R.Q.H. was supported by China Postdoctoral Science Foundation (Grant No. 2015T80069). Computational resources were provided by National Supercomputer Center in Guangzhou with Tianhe-2 Supercomputer and Physical Laboratory of High Performance Computing in Renmin University of China.
\end{acknowledgments}

\bibliography{ref}

\end{document}